# ADVANCED KELVIN PROBE OPERATIONAL METHODOLOGY FOR SPACE APPLICATIONS

G. CHIRITOI[1], E. M. POPESCU[1], A. A. RADU[1], A. CARAMETE[1], L. CARAMETE[1], A. PAVALAS[1], M. MARGARITESCU[2], A. E. ROLEA[2] T. NECSOIU[3], A. ENUICA[3]

[1] Institute of Space Science, 409 Atomistilor Street, P.O. Box MG-7, Magurele, Ilfov County, Romania, RO 077125
E-mail: *gabriel.chiritoi@spacescience.ro*
Email: *empopescu@spaescience.ro*

[2] National Institute for Research and Development in Mechatronics and Measurement Techniques, 6-8 Sos. Pantelimon, Sector 2, Bucharest, Romania

[3] S. C. Optoelectronica 2001 S. A., 409 Atomistilor Street, P.O. Box MG-22, Magurele, Ilfov County, Romania RO 077125

*Abstract*. We present a novel methodology for the operation of macroscopic Kelvin Probe Instruments. The methodology is based on the use of a harmonic backing potential signal to drive the tip-sample variable capacitance and on a Fourier representation of the tip current, allows for the operation of the instrument under full control and improves its scanning performance by a factor of 60 or more over that of currently available commercial instruments.

*Keywords*: Kelvin probe, instrumentation, space.

## 1. INTRODUCTION

Kelvin Probes (KPs) are instruments that allow for the direct non-contact measurement of the contact potential difference (CPD) between the surface of a sample and a metallic probe [1]. As the probe is made of physically and chemically stable materials (e.g. gold, or stainless steel) at the time scales relevant to the sample's surface and bulk materials interactions with the environment, the probe work function can be considered as a reference potential in the CPD measurement, and one can take the view that KP instruments are capable of measuring a sample's local work function, which can be assimilated to the local or average surface potential of the sample. We will adopt this latter view, and as such throughout the paper the terms "contact potential difference" and "surface potential" will be used interchangeably.

Traditionally, there are two major types of KP instruments, namely macroscopic KPs (MKPs) [2], [3], and microscopic KPs [4], [5], [6]. The former are generally instruments dedicated exclusively to the measurement of the surface potential of large surface area samples (typically $\geq$ 1 cm×1 cm), and in their scanning version they are quite slow, typically requiring more than 40 min to map a 10 cm×10 cm sample area with about 800-900 pixels resolution at high accuracy. The latter are Scanning Probe Microscope (SPM) or Atomic Force Microscope (AFM) instruments with surface potential measurement operation modes (or "options" in the dedicated SPM language) such as Scanning Capacitance Microscopy (SCM) or Electrostatic Force Microscopy (EFM) that can map the surface potential on very small areas of the sample surface (typically $\leq$ 200-300 $\mu m^2$). By contrast with the MKSs, SPM KP instruments are much faster, requiring no more than a couple of minutes to map a 100 $\mu m \times$ 100 $\mu m$ area of a sample with 256 × 256 pixels resolution.

However, when it comes to space applications such as the development of test masses and electrostatic suspension systems in experiments like SR-POEM [7] and LISA/NGO [8], or the study of spacecraft charging effects, one needs an instrument capable to accurately map the surface potential across large areas, with high enough resolution and fast enough to be compatible with the effects and phenomena that could degrade the performance of the experiments [9]-[13] or could have a negative impact on the spacecraft operation [14], [15]. Unfortunately, it becomes quite clear from the above considerations that neither MKP-type instruments (large scan area, but low pixel resolution and slow) nor SPM-type instruments (small scan area, but high pixel resolution and fast) can simultaneously satisfy such requirements, and that in fact for this kind of applications one would need to develop a more performant instrument combining the best operational characteristics of the two types of KP instruments. The development of such an instrument must necessarily rely on a thorough analysis of the construction and operation of the MKP and SPM-type instruments, and while such an analysis is beyond the scope of the present paper, at this point it will suffice to say that the most effective way to proceed about its development is to make key improvements on the performance of the MKP-type instruments.

Under these circumstances, the purpose of the present work is to present a novel methodology for the operation of an MKP instrument that will significantly improve its performance. The methodology relies on the use of a harmonic backing potential signal to drive the probe of an MKP instrument (as opposed to the switching backing potential current standard), and as it will be become clear in the remainder of the paper, this seemingly minor operational change will improve the instrument's scanning performance by a factor of more than 60.

The paper is organized as follows. In the Section 2 we give a brief review of the construction and operation of a standard MKP instrument, as well as of its performance limitations. In Section 3, we will present the novel operational methodology for an improved MKP instrument, and will discuss in detail its impact on the instrument performance. The paper will conclude in Section 3 with an overview of the results and with suggestions for further work.

## 2. STANDARD MKP OPERATION AND PERFORMANCE

As mentioned in the previous section, an MKP instrument is an instrument capable of measuring the surface potential of large area metallic samples. Its schematic construction is illustrated in Fig. 1 below. For more details, the reader referred to [1], [2] and [3].

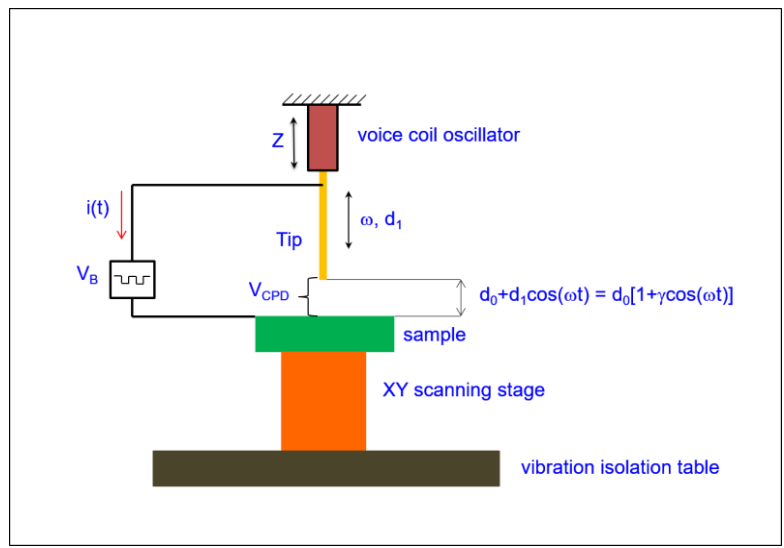

**Figure 1** – Schematic representation of an MKP instrument

A metallic sample is placed, in the scanning version of the instrument, on an XY scanning stage, which in its turn is placed on a vibration isolation table to attenuate environmentally induced oscillations. The KP probe consists of a metallic tip (usually gold or stainless still) attached to an electromechanical oscillator (essentially a voice coil oscillator) which oscillates it harmonically at a fixed (angular) frequency $\omega$ and amplitude $d_1$ in the Z direction while at the same time the XY stage scans it over the sample at a fixed tip-sample distance $d_0$ (measured in the absence of the tip oscillation). Assuming a flat tip, (e.g. a cylindrical metallic rod with one of its end faces parallel to the sample), the setup in Fig. 1 can be viewed as a variable parallel-plate tip-sample capacitor with harmonically varying inter-plate distance which is scanned over the sample surface.

In the standard MKP instrument configuration, this variable capacitor is polarized by a pulsed waveform voltage signal $V_B$ (a switching "backing potential" in the dedicated terminology) that will cause the flow of a current i(t) in the tip-sample electrical circuit. The measurement of this current, called the tip or probe current, allows one the tip-sample contact potential difference $V_{CPD}$, which, as explained earlier, can be assimilated to the local surface potential of the sample.

There are several methods for extracting $V_{CPD}$ from the measurements of the tip current i(t), but the one favored most by MKP manufacturers relies on the so-called quadratic fitting model (QFM) [16]. The method can be summarized as follows.

On each half-cycle of the $V_B$ square pulse waveform, the time-varying charge accumulated on the plates of the tip-sample capacitor is given by the expression:

$$Q(t) = (V_B - V_{CPD})C(t) \qquad (1)$$

where $V_B$ is the signed value of the backing potential amplitude, and C(t) is the time-varying tip-sample capacitance, and the corresponding tip current will be:

$$i(t) = \frac{dQ(t)}{dt} = (V_B - V_{CPD})\frac{dC(t)}{dt} \qquad (2)$$

The expression of the flat tip time-varying tip-sample capacitance is given by the standard expression:

$$C(t) = \frac{\varepsilon_0 A}{d(t)} \qquad (3)$$

where $\varepsilon_0$ is the free-space permittivity, A is the area of the tip end-face, and d(t) the instantaneous tip-sample separation. Using the notations in Fig.1, the latter will have the expression:

$$d(t) = d_0 + d_1 \cos(\omega t) \qquad (4)$$

where in eqn. (4) we have used for convenience a cosine function to describe the tip oscillation along the Z direction. It is moreover customary to rewrite eqn. (4) in the form:

$$d(t) = d_0[1 + \gamma \cos(\omega t)] \qquad (5)$$

where $\gamma = d_1/d_0$ is called the tip (amplitude) modulation index, and with eqns. (3) and (5), the expression of the tip current will take the form:

$$i(t) = \omega C_0 (V_B - V_{CPD}) \frac{\gamma \sin(\omega t)}{[1 + \gamma \cos(\omega t)]^2} \qquad (6)$$

with $C_0 = \varepsilon_0 A/d_0$ the average tip-sample capacitance. A simulation of $V_B$ and of the tip signal given by eqn. (6) for a 3 mm diameter tip, $d_0$ = 50 μm, $\gamma$ = 0.9, $V_B$ = ± 1 V, $V_{CPD}$ = 0.5 V, and typical values for the tip oscillation frequency (100 Hz) and $V_B$ pulse width (25 Hz) is shown in Fig. 2 below. For reasons of clarity, the amplitudes of the signals have been rescaled by convenient arbitrary factors.

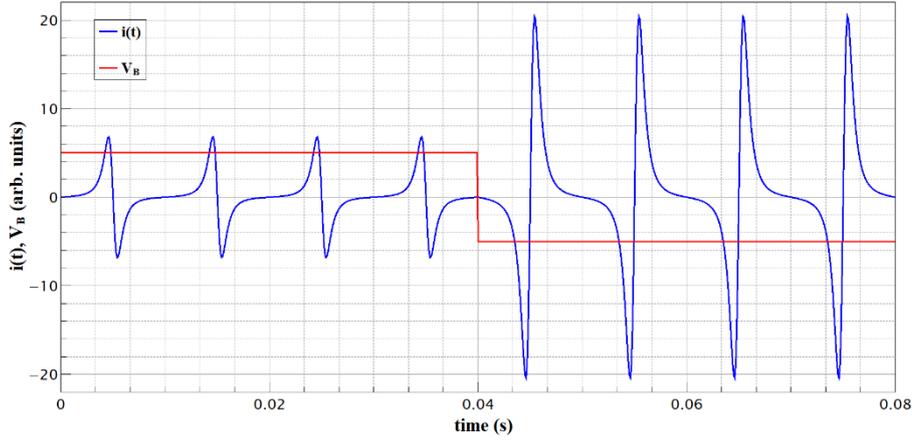

**Figure 2** – Simulation of the MKP tip signal (blue) with switching backing potential (red)

The QFM operational method relies on the existence of several local extrema of the tip signal and on their symmetry in a given $V_B$ half-cycle, as illustrated in Fig. 2. The values of the local extrema are determined using a quadratic polynomial least-squares fitting procedure for the data in the neighborhood of each such local extremum, and then, using the symmetry of the tip signal with respect to the time axis, they are used to define a tip signal amplitude for a given $V_B$ half-cycle (e.g. as the average of the magnitude of the semi-difference of two successive local extrema values over all available extrema in the respective half-cycle). Since according to eqn. (6) the amplitude of the tip current signal is a linear function of ($V_B$-$V_{CPD}$), once the tip signal amplitudes have been determined for two successive $V_B$ half-cycles (i.e. for two $V_B$ values), they are interpolated with a linear function of $V_B$ which is then used to determine $V_{CPD}$ as the value of $V_B$ corresponding to the vanishing of the tip signal amplitude. The procedure is then repeated for each $V_B$ half-cycle for as long as the instrument is operation.

Insofar, the MKP operation as described above seems simple enough and fast enough to allow an MKP instrument to operate at $V_{CPD}$ measurement rates limited solely by the performance of its data acquisition system. The reality, however, is a little bit more complicated, as illustrated in Fig. 3 where we show the preamplifier output raw tip signal of an MTS KP 6500 instrument [2] measuring the contact potential difference of a copper sample with a 3 mm diameter flat tip in a non-scanning (static) configuration with $V_B = \pm 2V$ and a tip oscillation frequency of 150.1 Hz.

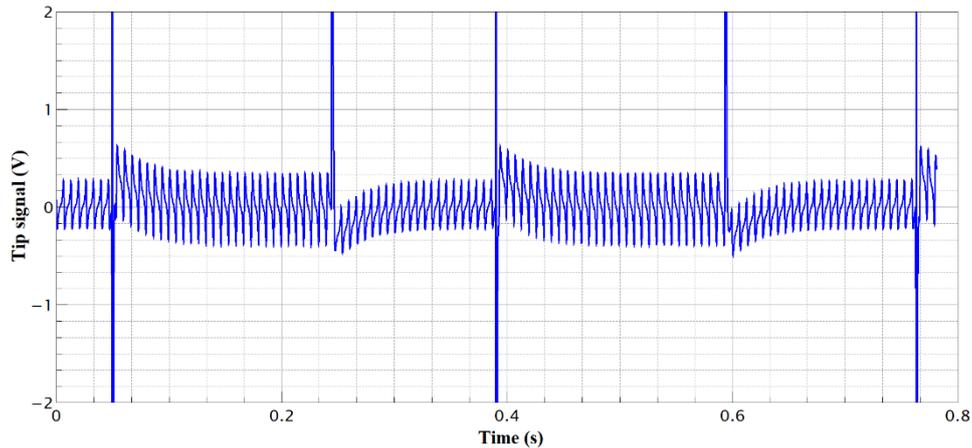

**Figure 3** – Real MKP tip current data for the static measurement of the surface potential of a copper test sample with a stainless steel tip

By simple inspection of Fig. 3, one can immediately notice that there are significant differences between the real tip signal and the simulated one in Fig. 2, despite of the fact that their shape over a tip oscillation period are very similar. Indeed, the tip signal in Fig. 3 shows large overshoots at the switching of the backing potential $V_B$, followed by visible relaxation effects which prevent the use of the QFM method – and hence the measurement of the sample's surface potential – before the stabilization of the signal which occurs some 60-70 ms later. In fact, MKP instruments have built in $V_{CPD}$ measurement delays (user controllable in the case of the KP 6500) designed to prevent the erroneous measurement of $V_{CPD}$ due to these overshooting and relaxation effects.

This means that in reality, the $V_{CPD}$ measurement rate of a real MKP instrument is not limited by the performance of its data acquisition system, but rather by the switching backing potential used to polarize the tip-sample variable capacitor which induces undesired effects in the tip signal detection electronics. Under these circumstances, MKP instruments have typical tip current amplitude and $V_{CPD}$ measurement rates of the order of hundreds of milliseconds (~ 400 ms/2.5 Hz in the example shown in Fig. 3), which are rather incompatible with the space applications and studies envisioned for their use.

If we now take into consideration scanning, things become even more complicated. Scanning a large sample requires three major ingredients in addition to a fast measuring (static) instrument, namely a fast scanning XY stage, full control of the instrument during the scanning process, and a scanning procedure. Fast and precise scanning stages with large travel are not very hard to come by at a price [17], so the issues left are the full control of the instrument during the scanning process and the scanning procedure.

For an MKP instrument, full control of the instrument means that instrument should be able to keep constant during scanning and $V_{CPD}$ measurements both the average tip-sample separation distance $d_0$ and the tip oscillation amplitude $d_1$ by means of appropriate active (feedback) controls. The control of the tip-sample spacing is not hard to implement in the switching backing potential QFM operational method, since according to eqn. (6) the slope of the tip signal amplitude vs. $V_B$ interpolation line mentioned earlier is proportional to $\gamma/d_0 \sim 1/d_0^2$. As such, MKP instruments have active controls that allow for the operation of the instrument at constant slope, and hence at constant tip-sample separation distance $d_0$. At the practical level, the active control may be implemented by varying DC offset current of the tip oscillation voice coil driver (e.g. in the case of the MTS KP 6500 instrument) or by using an additional Z-stage for the probe (e.g. in the case of the KP Technology SKP5050 instrument [3]). However, it should be noted that the above active control for $d_0$ is very slow, the tip position update rate being no faster than the $V_{CPD}$ measurement rate which, as mentioned earlier, is of the order of several Hz at best.

The situation is entirely different when it comes to controlling the tip oscillation amplitude $d_1$. Its control means in fact the precise control of the AC signal applied to the tip oscillation voice coil driver, otherwise, as the voice coil heats up during the instrument operation, it will cause the tip oscillation amplitude to change in time. This can lead at best to errors in the $V_{CPD}$ measurement or the instrument losing the measurement signal, and at worst to the tip actually ramming the sample. Rather unfortunately, the switching backing potential QFM operational method does not offer an obvious way to control the tip oscillation amplitude, and while it is possible to do so [16], to the best knowledge of the present authors, no commercially available MKP instrument has incorporated this type of control in its operation.

Referring now to the scanning procedure, due to the slow $d_0$ control and to the fact that MKPs can operate at values of $d_0$ as low as 10 μm and of the modulation index $\gamma$ as high as 0.95, the scanning procedure of choice for commercially available MKP instruments is of the type *move-approach-measure-raise-move* to prevent the tip to ram into the sample during its translation above the sample. In this procedure, the tip is moved by the XY stage at a safe height above the sample to the location on where a measurement has to be taken, an automatic approach is done by the instrument to lower the tip at the preset $d_0$ setpoint, $V_{CPD}$ is measured under active $d_0$ control, the tip is raised back at a safe height above the sample and the XY stage moves the tip to the next location for a new $V_{CPD}$ measurement.

Assuming a tip current amplitude measurement rate of about 400 ms like in the previous practical example, at least 3 time steps of the same magnitude for the tip approach to the sample with one tip current amplitude measurement per step (the first two steps are used to give a first measurement of $d_0$ with respect to the setpoint, and the last two to verify that the active control has performed the necessary correction and the instrument is actually operating close to the preset setpoint) and at least one more amplitude measurement to determine $V_{CPD}$, it all adds up to at least 1.6 s that are necessary for a single $V_{CPD}$ measurement at each point during scanning. Adding to this about 1 s for the translation of the tip between two $V_{CPD}$ measurement points on the sample, one gets an effective $V_{CPD}$ measurement rate during scanning of about 3 s/0.33 Hz, which easily amounts to a total time of more than 40 min for scanning a 10 cm×10 cm sample area at 800-900 pixels resolution, as stated in the previous section.

This example concludes our description of the switching backing potential MKP instrument operation. In the next section, we will present our approach to improving the performance of MKP instruments. This approach relies on the replacement of the switching backing potential signal driving of the tip with a harmonic one, and on the corresponding replacement of the QFM method for the $V_{CPD}$ measurement with a more appropriate one using the Fourier frequency content of the tip signal and allowing for the full control of the instrument during scanning.

### 3. A NOVEL METHOD FOR FAST MKP INSTRUMENT OPERATION

As stated above, our approach for improving the performance of MKP instruments relies essentially on the use of a harmonic backing potential $V_B$ to polarize the tip-sample capacitor, and on a method for the determination of the contact potential difference $V_{CPD}$ that makes use in a natural way of Fourier frequency content of the resulting tip-sample content. This method can be summarized as follows.

For a harmonic backing potential of the form:

$$v_B(t) = V_{B0} \sin(\omega_B t) \qquad (7)$$

with $V_{B0}$ the amplitude of the backing potential $v_B(t)$ whose harmonic time variation has been conveniently chosen to be sinusoidal with (angular) frequency $\omega_B$, eqn. (1) takes the form:

$$Q(t) = [V_{B0}\sin(\omega_B t) - V_{CPD}]C(t) \tag{8}$$

and the tip current will be given by the expression:

$$i(t) = \frac{dQ(t)}{dt} = [V_{B0}\sin(\omega_B t) - V_{CPD}]\frac{dC(t)}{dt} + \omega_B C(t)V_{B0}\cos(\omega_B t) \tag{9}$$

Using eqns. (4)-(6), the expression of the tip current in eqn. (9) can be put into the form:

$$i(t) = \omega_C C_0 [V_{B0}\sin(\omega_B t) - V_{CPD}]\frac{\gamma\sin(\omega_C t)}{[1+\gamma\cos(\omega_C t)]^2} + \frac{\omega_B C_0 V_{B0}\cos(\omega_B t)}{[1+\gamma\cos(\omega_C t)]} \tag{10}$$

with $\omega_C$ the tip oscillation frequency and $\gamma$, $C_0$ as defined in Section 2.

In practice, as most commercially available data acquisition boards still have a single clock for the timing of the input and output channels, it is convenient that the frequencies $\omega_B$ and $\omega_C$ be one an integer multiple of the other, e.g.:

$$\omega_C = k_0 \omega_B;\ k_0 = 0, 1, 2, \ldots \tag{11}$$

As mentioned earlier, our method of choice for the determination of the contact potential difference $V_{CPD}$ and of the instrument full control parameters $d_0$, $\gamma$ from the tip signal described by eqn. (10) relies on the exploitation of the tip current signal frequency content in a least-squares fitting procedure. For this purpose, it is convenient to use the following Fourier series expansions of the trigonometric functions of argument $\omega_C t$ in the RHS of eqn. (10) [16]:

$$\frac{1}{[1+\gamma\cos(\omega_C t)]} = \frac{2}{\sqrt{1-\gamma^2}}\left\{1 + \sum_{n=1}^{\infty}(-1)^n\left[\frac{\gamma}{1+\sqrt{1-\gamma^2}}\right]^n\cos(n\omega_C t)\right\} \tag{12}$$

$$\frac{\gamma\sin(\omega_C t)}{[1+\gamma\cos(\omega_C t)]^2} = \frac{2}{\sqrt{1-\gamma^2}}\left\{1 + \sum_{n=1}^{\infty}(-1)^{n-1}n\left[\frac{\gamma}{1+\sqrt{1-\gamma^2}}\right]^n\sin(n\omega_C t)\right\} \tag{13}$$

By using eqns. (12) and (13) with a choice of $k_0 = 3$ to optimize the Fourier frequency content of the signal an straightforward trigonometric and algebraic manipulations, eqn. (10) can be recast in the form:

$$i(t) = \frac{1}{2}BV_{B0}\cos(\omega_B t) + \frac{1}{2}ABV_{B0}\cos(2\omega_B t) - \frac{3}{2}ABV_{CPD}\sin(3\omega_B t)$$
$$- ABV_{B0}\cos(4\omega_B t) - \frac{5}{4}A^2 BV_{B0}\cos(5\omega_B t)$$
$$+ 3A^2 BV_{CPD}\sin(6\omega_B t) + \frac{7}{4}A^2 BV_{B0}\cos(7\omega_B t)$$
$$+ 2A^3 BV_{B0}\cos(8\omega_B t) - \frac{9}{2}A^3 BV_{CPD}\sin(9\omega_B t)$$
$$- \frac{5}{2}A^3 BV_{B0}\cos(10\omega_B t) - \frac{11}{4}A^4 BV_{B0}\cos(11\omega_B t) \quad (14)$$
$$+ 6A^4 BV_{CPD}\sin(12\omega_B t) + \frac{13}{4}A^4 BV_{B0}\cos(13\omega_B t)$$
$$+ \frac{15}{4}A^5 BV_{B0}\cos(15\omega_B t) + \cdots$$
$$\equiv \sum_{n=1}^{\infty} i_{Sn}\sin(n\omega_B t) + \sum_{n=1}^{\infty} i_{Cn}\cos(n\omega_B t)$$

where in eqn. (14) we have used for simplicity $i_{Sn}$, $i_{Cn}$ as the generic coefficients of the tip current Fourier frequency components, as well the following notations:

$$A = \frac{\gamma}{1+\sqrt{1-\gamma^2}} \quad (15)$$

$$B = C_0 \omega_B \quad (16)$$

It is clear from eqn. (14) that $V_{CPD}$ as well as the full control parameters $d_0$, $\gamma$ can be determined by using the coefficients of just the first three Fourier frequency components, in which case we have:

$$V_{CPD} = -\frac{1}{3}V_{B0}\left(\frac{i_{S3}}{i_{C2}}\right) \quad (17)$$

$$d_0 = -\frac{1}{2}\varepsilon_0 A \omega_B \left(\frac{i_{S1}}{i_{C1}}\right) \quad (18)$$

$$d_1 = \frac{2d_0}{\left[\left(\frac{i_{C2}}{i_{C1}}\right) + \left(\frac{i_{C1}}{i_{C2}}\right)\right]} \quad (19)$$

In practice, the Fourier coefficients necessary for the determination of the $V_{CPD}$, $d_0$, and $\gamma$ model parameters, and in fact all the amplitudes of the Fourier components in the RHS of eqn. (14), can be determined from a least-squares fitting of the tip current measured data with the model given by eqn. (14). For an appropriate choice of partition for the period $T_B=2\pi/\omega_B$, i.e. of the data acquisition rate, it comes as no surprise that the least-squares estimation of the Fourier amplitudes yields the rather trivial relations:

$$i_{Sn} = \frac{\omega_B}{\pi} \sum_{\text{data points/period}} i_r \sin(n\omega_B t_r) \tag{20}$$

$$i_{Cn} = \frac{\omega_B}{\pi} \sum_{\text{data points/period}} i_r \cos(n\omega_B t_r) \tag{21}$$

which are nothing more than the discrete counterparts of the standard integral Fourier relations for the calculation of a Fourier series coefficients.

It should be noted that in fact one can use for the least square estimation of $V_{CPD}$, $d_0$, and $\gamma$ model parameters a partial Fourier model (PFM) [16] consisting of the truncation of the Fourier series in the RHS of eqn. (14) to a convenient number of terms without changing in any way the content of eqns. (17)-(21). In the present case, the number of terms can be as little as four, and hence the PFM will only contain just the first three Fourier frequency components necessary for the model parameter estimation. For reasons of simplicity, in all of the following we will limit ourselves to just this minimal PFM, as its use does not affect in any way the generality of our arguments and conclusions.

With these considerations, we are now ready to proceed with the analysis of the harmonic backing potential method by comparing it to the switching backing potential method of the previous section. For this purpose, and using the same parameters as for the signal depicted in Fig. 2, the simulation of the tip signal given by eqn. (10) is illustrated in Fig. 4 below. For reasons of clarity, the amplitudes of the signals have been rescaled by the same convenient arbitrary factors used in Fig.2, and a zero phase-shift was assumed between the backing potential and tip current signals.

Referring now to the tip current signal depicted in Fig. 4, several important comments are in order at this time. First of all, driving the tip with a pure AC (harmonic) backing potential signal $v_B(t)$ removes the stringent need for MKP instruments to have ultrafast fast pulse detection electronics capable to respond properly to a switching backing potential signal. In other words, using a purely AC backing potential tip driving signal eliminates from the start the tip current overshooting and relaxation effects visible in Fig. 3, and correspondingly the delay times in the $V_{CPD}$, $d_0$, and $d_1$ measurements which, as shown in Fig. 3, can be of the order of 60-70 ms and which adds significantly to the total scanning time.

Secondly, once the need for ultrafast pulse detection electronics has been eliminated, one can easily operate an MKP instrument with harmonic backing potential under full control at much larger frequencies than the current standard couple of hundreds of Hz. Indeed, the voice coil can operate without any problem up to frequencies of the order of 20 kHz, and one can use off-the-shelf electronic components for the construction of tip current preamplifiers (e.g. AD713JN operational amplifiers [18]) that can operate without any problem at the 60 kHz bandwidths imposed by the constraint in eqn. (11) and by the use of a minimal PFM.

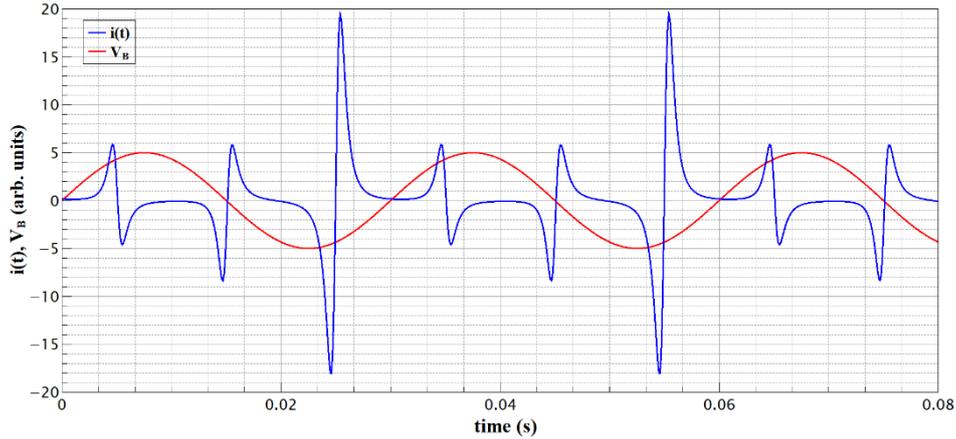

**Figure 4** – Simulation of the MKP tip signal (blue) with harmonic backing potential (red)

Under these circumstances, by taking rather conservative values for the operational frequencies of a harmonic backing potential MKP, e.g. a backing potential frequency of say 10 KHz and a tip oscillation frequency of say 30 kHz, if one takes the view of standard switching potential MKP instruments that the $V_{CPD}$, $d_0$, and $d_1$ measurement rate is determined by the frequency of the backing potential, the time required by a harmonic backing potential MKP instrument to perform one measurement of $V_{CPD}$, $d_0$, and $d_1$ will be 100 μs plus the data processing time and control parameters update time at the hardware level which is of the order of a few tens of μs. This implies a rather effortless decrease in the time required for the measurement of $V_{CPD}$, $d_0$, and $d_1$ of more than three orders of magnitude from ~ 400 ms in the case of the switching potential MKP to ~ 150 μs for the harmonically driven MKP.

Moreover, if one adopts a much larger view of the data acquisition process underlying the estimation of $V_{CPD}$, $d_0$, and $d_1$ as model parameters according to the methodology put forward in eqns. (17)-(21), the parameters' measurement rate could be pushed in principle up to the data instrument's data acquisition rate. Indeed, one can use a FIFO sliding sum procedure for the calculation of the Fourier coefficients in eqns. (21)-(22) each time a new tip current signal measurement is performed by the instrument, and update accordingly the estimations of the model parameters

$V_{CPD}$, $d_0$, and $d_1$ at the same rate. Of course, in practice one can never reach the instrument's data acquisition rate for a variety of reasons including the data processing time and the PFM control parameters' update time at hardware level. Nevertheless, and once more without much effort, using such a FIFO sliding sum method for a convenient number of newly acquired data points, e.g. during one period $T_C=2\pi/\omega_C$ of the tip oscillation, the model parameters' measurement time can be further decreased by a factor of about two, from ~150 μs to ~ 80 μs.

Finally, and referring now to scanning, the method we propose allows for much higher scanning speeds and for the full instrument control during scanning. Indeed, by eliminating the switching backing potential driving of the tip-sample capacitor one automatically eliminates the need for a *move-approach-measure-raise-move* scanning procedure and allows the instrument to use faster scanning procedures such as continuous scanning with measurements on the fly in which the tip is moved at constant speed along a scan line under full instrument control and $V_{CPD}$ data values are recorded at given time intervals corresponding to the user-set number of pixels per scan line. In this case, and ignoring the acceleration and deceleration times at the ends of the scan line, the maximum scanning speed is determined by the data measurement and processing total time ensuring the scanning under full instrument control. As mentioned above, a reasonable value of this time is ~80 μs, which means in order to scan a line with the instrument under a reasonably fast instrument control one would need to have a displacement of a 3 mm diameter tip by at most one hundredth of its diameter during two updates of the control parameters $d_0$ and $d_1$, i.e. a scanning speed of ~375 mm/s, which is consistent with the performance of commercially available stepper motor stages [17]. Under these circumstances, and by adding the acceleration and deceleration times at the ends of the scan line, a 10 cm long scan line would take ~ 0.7 s to complete. Correspondingly a 10 cm×10 cm scan with a 30×30 pixel resolution will take, including the translations in the slow scan direction, less than 40 seconds to complete under full instrument control, which is more than 60 times faster than the 40 min or more it takes a switching backing potential MKP instrument to perform the same scan.

## 4. SUMMARY AND CONCLUSIONS

To summarize, the goal of the present work is the presentation of a novel operational methodology for MKP instruments that can significantly improve their performance over that of currently available commercial instruments.

For this purpose, we have provided a detailed description of the operational methodology of commercial MKP instruments, with emphasis the causes limiting their performance. As such we have shown that the main sources of limitations in the measurement performance of switching backing potential MKP instrument is the switching backing potential itself, coupled with a rather cumbersome QFM procedure for the determination of the contact potential difference $V_{CPD}$ and tip-sample separation distance $d_0$ and with a very slow scanning procedure. All these three sources combined together yield, besides just a partial control of the instrument during its operation, to large effective $V_{CPD}$ and $d_0$ measurement rates of ~0.33 Hz, leading ultimately to scanning times of more than 40 min scan times for large samples at 800-900 pixels resolution which is the standard in the field.

The solution proposed to overcome these limitations relies on the replacement of the switching backing potential signal driving the tip-sample variable capacitor with a harmonic one, and on a partial Fourier series representation of the tip current which allows for the determination of $V_{CPD}$ and of both instrument control parameters $d_0$ (the tip-sample separation distance) and $d_1$ (the tip oscillation amplitude) by means of a least-squares optimal estimator. This methodology eliminates from the start the large delays associated with the tip current measurements, and together with a conveniently chosen sliding sum procedure for the estimation of the coefficients in the Fourier series representation of the tip current, it allows for the increase in the $V_{CPD}$, $d_0$, and $d_1$ effective measurement rates up to more than 12 kHz. Moreover, the combination of this methodology with commercially available fast XY scanning stages yields scan times of ~ 40 seconds for the same large sample at 800-900 pixels resolution.

As such, we consider that the operational methodology proposed here is very promising for improving the performance of MKP instruments and worth to be implemented in practice. In fact, an instrument using this methodology is currently under development at the Institute of Space Science, and we will report on its operational performance in a future publication.

AKNOWLEDGEMENTS

The present work has been supported by MEN/MCI-UEFISCDI through the Priority Domains Partnerships Program, contract 279/2014.

**REFERENCES**

1. I. D. Baikie, *Old Principles . . . New Techniques: A Novel UHV Kelvin Probe and its Application in the Study of Semiconductor Surfaces,* Enschede, The Netherlands (1988)
2. McAllister Technical Services KP 6500: https://mcallister.com/product/uhv-kelvin-probe-system/
3. KP Technology SKP 5050: http://www.kelvinprobe.com/
4. Park Systems: http://www.parkafm.com/index.php/park-spm-modes/electrical-properties/235-scanning-capacitance-microscopy-scm
5. NT-MDT Spectrum Instruments: http://www.ntmdt-si.com/spm-principles/view/scanning-capacitance-microscopy .
6. Nanosurf: https://www.nanosurf.com/en/application/523-electrostatic-force-microscopy-efm
7. R. D. Reasenberg et al, *Class. Quantum Grav.* **28**, 094014 (2011)
8. K. Danzmann, *Class. Quantum Grav.* **13**, A247 (1996)
9. C. C. Speake, Class. Quantum Grav. **13**, A291 (1996)
10. N. A. Robertson et al, Class. Quantum Grav. **23**, 2665 (2006)
11. M. Araujo et al, *Astropart. Phys.* **22**(5-6), 451 (2005)
12. D. N. A. Shaul et all*, Class. Quant. Grav.* **22**(10), S297 (2005)
13. M. Armano et al, *Phys. Rev. Lett.* **118**, 171101 (2017)
14. H. B. Garrett, *Rev. Geophys. Space Phys.* **19**(4), 577 (1981)
15. X. Meng et al, *IEEE Trans. Plasma Sci.* **PP**(9), 1 (2017)
16. E. M. Popescu, *Class. Quant. Grav.* **28**, 225005 (2011)
17. Newport ONE-XY200 motorized XY stage: https://www.newport.com/p/ONE-XY200
18. AD713JN Operational Amplifier: http://www.analog.com/media/en/technical-documentation/data-sheets/AD713.pdf